\documentclass[conference]{IEEEtran}
\IEEEoverridecommandlockouts
\usepackage{cite}
\usepackage{amsmath,amssymb,amsfonts}
\usepackage{algorithmic}
\usepackage{graphicx}
\usepackage{textcomp}
\usepackage{xcolor}
\usepackage{hyperref}
\usepackage{multirow}
\def\BibTeX{{\rm B\kern-.05em{\sc i\kern-.025em b}\kern-.08em
    T\kern-.1667em\lower.7ex\hbox{E}\kern-.125emX}}
\begin{document}

\title{Component-Level Ensemble Fusion for Speech and Environmental Sound Deepfake Detection}

%\author{André Runewicz, Karla Schaefer, Martin Steinebach}

\author{
\IEEEauthorblockN{Andr\'e Runewicz, Karla Schäfer, Martin Steinebach}
\IEEEauthorblockA{
\textit{Fraunhofer SIT | ATHENE – National Research Center for Applied Cybersecurity}\\
Darmstadt, Germany\\
\{andre.runewicz, karla.schaefer, martin.steinebach\}@sit.fraunhofer.de
}
}

\maketitle

\begin{abstract}
This paper describes our submission to the ICME 2026 ESDD2 challenge on
environment-aware speech and sound deepfake detection. The task requires
five-class classification of audio clips in which speech, environmental
sound, both components, or neither component may be spoofed. We propose a
component-level ensemble system based on four publicly available
pre-trained anti-spoofing models: XLSR-Mamba, DF-Arena, SLS, and
TCM-ADD. Each model is fine-tuned on the official CompSpoofV2
development data using three binary heads for original, speech, and
environmental sound detection. We further train RawBoost-augmented variants and combine selected checkpoints using margin-space score fusion. A component-wise fusion strategy with lightweight head- and class-bias
calibration yields our best configuration, reaching 0.7715 macro-F1 on the
evaluation set and 0.7828 macro-F1 on the test set, ranking 5th out of 31 teams in the final ranking phase and substantially outperforming the official baseline.
\end{abstract}


\section{Introduction}
\label{sec:intro}
Audio deepfakes pose an increasing threat to the integrity of digital
communication. Recent advances in speech synthesis and voice conversion
have made it increasingly easy to generate convincing synthetic speech
from only a few seconds of reference audio. For example, VALL-E-X can
generate cross-lingual speech from a short 3--10 second recording
\cite{zhang2023speak}. This low barrier to entry increases the risk of
misuse in fraud, impersonation, and disinformation.

Most benchmark-driven audio deepfake detection research has focused on
deciding whether an entire recording is bona fide or spoofed. However, real-world audio is often compositional, typically containing
foreground speech and background environmental sound. These components can be manipulated independently. For instance, an attacker may replace the speech while preserving the original acoustic background, or insert bona fide speech into a synthetic or manipulated environment. Such component-level manipulations are more challenging than conventional utterance-level
spoofing, because an unmodified component may mask artifacts in the
manipulated one.

The ESDD2: Environment-Aware Speech and Sound Deepfake Detection
challenge addresses this setting by requiring systems to classify audio
clips according to the authenticity of both the speech and environmental
sound components \cite{zhang2026esdd2}. The challenge is based on the CompSpoofV2 dataset \cite{zhang2025compspoof,zhang2025esdd2compspoofv2}, whose evaluation and test splits contain unseen generated audio and differ in distribution from the training and validation splits, making robust generalization particularly important.

In this paper, we describe our submission to the ESDD2 challenge. Our
system adapts four publicly available pre-trained anti-spoofing models
to the component-level setting by adding three binary heads for
original, speech, and environmental sound detection. We train both
standard and RawBoost-augmented variants and combine selected
checkpoints using margin-space ensemble fusion. Finally, we apply
lightweight head- and class-bias calibration to improve the five-class
decision. The resulting system ranks 5th in the final ranking phase and
3rd in the preparation phase, substantially outperforming the official
baseline.

The remainder of this paper is organized as follows:
Section~\ref{sec:rw} reviews related work, Section~\ref{sec:em}
describes the proposed system, Section~\ref{sec:results} presents the
experimental results, and Section~\ref{sec:conclusion} concludes the
paper.

\section{Related Work}
\label{sec:rw}
Audio deepfake detection has traditionally been formulated as a binary
anti-spoofing task, where each recording is classified as bona fide or
spoofed \cite{9358099,yamagishi2021asvspoof,wang24_asvspoof}. Early systems relied on handcrafted acoustic features, such as
linear frequency cepstral coefficients (LFCCs), followed by conventional
classifiers such as support vector machines \cite{hua2021towards}.
Subsequent work introduced deep neural architectures operating on
time-frequency representations, including mel-spectrograms, as well as
end-to-end waveform-based models such as RawNet2 \cite{tak2021end}.
These approaches improved performance on controlled benchmarks, but
their robustness under distribution shift remains a major challenge.

Generalization to unseen attacks and real-world recordings is one of the
central problems in audio deepfake detection. Müller et al.
\cite{muller2022does} showed that detectors performing well on
ASVspoof 2019 evaluation data can degrade substantially when evaluated
on in-the-wild recordings. This observation is particularly relevant for
ESDD2 \cite{zhang2025esdd2compspoofv2}, where the training/validation and evaluation/test splits follow different source distributions and the leaderboard splits include unseen generated audio.

Recent state-of-the-art detectors increasingly rely on self-supervised
learning (SSL) representations. Models based on Wav2Vec 2.0
\cite{baevski2020wav2vec}, HuBERT \cite{hsu2021hubert}, and XLS-R \cite{babu22_interspeech} have been used as strong front-ends for downstream spoofing detection
\cite{tak2022automatic,xiao2025xlsr,zhang2024audio,truong24b_interspeech}.
The motivation is that SSL models trained on large and diverse speech
corpora can provide robust acoustic representations that transfer better
to unseen spoofing attacks than task-specific handcrafted features. The
base detectors used in our system follow this direction and build on
pre-trained SSL backbones.

Beyond speech spoofing, recent work has started to consider
manipulations of environmental or background audio \cite{yi2024scenefake, cheng2024envfake, ouajdi2024detection, xie2024fakesound, kim2019audiocaps, yin2025envsdd}. This direction is important because real recordings contain both foreground speech and background acoustic scenes, and either component can be modified independently. Environmental sound manipulations pose distinct challenges, since background sounds differ from speech in their spectro-temporal structure. Prior work has shown that even subtle modifications to the background component of an audio signal can degrade the performance of existing detectors \cite{shih2024does,schafer2025audio}. This motivates component-level detection methods that explicitly model both speech and environmental sound authenticity.

Since the individual detectors considered in this work exhibit different
strengths across component heads, our system uses component-wise
score-level fusion rather than a single global fusion rule. This allows
the original, speech, and environmental sound decisions to rely on
different subsets of models.

\section{Experimental Methodology}
\label{sec:em}
This section describes our component-level detection pipeline, including
the task formulation, dataset protocol, base detector fine-tuning,
RawBoost augmentation, score export, ensemble fusion, and post-hoc
calibration.

\subsection{Task Formulation}
The ESDD2 task is a five-class component-level audio deepfake
detection problem. Given an input audio clip \(x\), the system must predict one of five labels: original audio, bona fide speech with bona fide environment, spoofed speech with bona fide environment, bona fide speech with spoofed environment, or spoofed speech with spoofed environment. Instead of training a single five-class classifier, we factorize the task into three binary decisions that follow the structure of the challenge: whether the clip is original or mixed, whether the speech component is bona fide or spoofed, and whether the environmental component is bona fide or spoofed.

For each input waveform, the system estimates three component-level
quantities:
\begin{equation}
\begin{aligned}
q_o &= P(O=1 \mid x), \\
q_s &= P(S=1 \mid O=0,x), \\
q_e &= P(E=1 \mid O=0,x),
\end{aligned}
\end{equation}
where \(O=1\) denotes original audio, \(O=0\) denotes mixed audio, and
\(S=1\) and \(E=1\) denote bona fide speech and bona fide environment,
respectively. The speech and environment heads are therefore interpreted
inside the mixed-audio branch. We use the factorized approximation
\(P(S,E\mid O=0,x)\approx P(S\mid O=0,x)P(E\mid O=0,x)\), which gives
\begin{equation}
\begin{aligned}
p_0 &= q_o, \\
p_1 &= (1-q_o) q_s q_e, \\
p_2 &= (1-q_o)(1-q_s) q_e, \\
p_3 &= (1-q_o) q_s (1-q_e), \\
p_4 &= (1-q_o)(1-q_s)(1-q_e).
\end{aligned}
\end{equation}
Here, class \(0\) denotes original audio, while classes \(1\)--\(4\)
correspond to the four possible bona fide/spoofed combinations of the
speech and environmental components. This factorized formulation allows
the system to model the original/mixed decision separately from the two
component-level authenticity decisions.
This is a decision-level factorization, not a claim of strict
statistical independence. A direct five-class head was also evaluated
during development, but performed slightly worse and was not used in the
final system.

\subsection{Dataset and Challenge Protocol}
We use the official CompSpoofV2 dataset released for the ESDD2 challenge\cite{zhang2025esdd2compspoofv2,zhang2026esdd2}. CompSpoofV2 contains more than 250k four-second audio clips, corresponding to approximately 283 hours of audio. The dataset is divided into training, validation, evaluation, and test splits, as summarized in Table~\ref{tab:data_splits}. The training and validation splits share the same source distribution, whereas the evaluation and test splits share a different distribution and include unseen generated audio. The class distribution is moderately imbalanced across splits. %This distributional mismatch is central to the challenge setup and makes validation-set model selection less reliable than in a conventional in-distribution setting.

\begin{table}[t]
\centering
\caption{CompSpoofV2 data splits used in the ESDD2 challenge.}
\label{tab:data_splits}
\begin{tabular}{lrrl}
\hline
Split & \# clips & Labels & Use in our system \\
\hline
Train & 175,361 & yes & optimization \\
Validation & 24,864 & yes & validation monitoring \\
Evaluation & 27,605 & hidden & preparation-phase submissions \\
Test & 27,603 & hidden & final-phase submissions \\
\hline
\end{tabular}
\end{table}

All systems are trained using only the official development data. The
training split is used for parameter optimization and the validation
split is used for monitoring training progress. No evaluation or test
samples are used for gradient-based training, and no additional audio
datasets are used for fine-tuning or calibration. The only external
resources used by the system are publicly available pre-trained models
released before January 2026, which are disclosed in
Section~\ref{subsec:base_detectors}.

The challenge consists of a preparation phase and a final ranking phase.
During the preparation phase, participants submit predictions for the
evaluation set and receive leaderboard feedback. During the final ranking
phase, participants submit predictions for the test set. Because the
validation split follows the training distribution while the evaluation
and test splits follow a different distribution, the validation-optimal
checkpoint was not always the best checkpoint on the leaderboard. We
therefore export candidate checkpoints after each training epoch and
use the preparation-phase leaderboard feedback for checkpoint and fusion
selection. The selected checkpoints and predefined fusion recipes are
then applied to the test set during the final phase.

\subsection{Base Detectors}\label{subsec:base_detectors}
We fine-tune four publicly available pre-trained anti-spoofing models:
XLSR-Mamba \cite{xiao2025xlsr}, DF-Arena 1B \cite{kulkarni2026compactsslbackbonesmatter}, SLS \cite{zhang2024audio}, and TCM-ADD \cite{truong24b_interspeech}. Table~\ref{tab:base_detectors} summarizes the four pre-trained detectors used as backbones. We denote these systems as \(A\), \(B\), \(C\), and \(D\), respectively. For each architecture, we train one model without data augmentation and one model with RawBoost augmentation \cite{tak2022rawboostrawdataboosting}. We use the subscript ``RB'' to denote RawBoost variants, e.g., \(A_{\mathrm{RB}}\).

\begin{table}[t]
\centering
\caption{Pre-trained base detectors used in the ensemble.}
\label{tab:base_detectors}
\setlength{\tabcolsep}{3pt}
\small
\begin{tabular}{llll}
\hline
ID & Model & Front-end & Backend \\
\hline
$A$ & XLSR-Mamba & XLSR2-300M & Bi-Mamba + attention pooling \\
$B$ & DF-Arena & XLS-R 1B & layer weighting + Conformer \\
$C$ & SLS & XLSR2-300M & layer weighting + MLP \\
$D$ & TCM-ADD & XLSR2-300M & TCM-enhanced Conformer \\
\hline
\end{tabular}
\end{table}

In addition to the main DF-Arena run \(B\), we train one additional
DF-Arena model without RawBoost, denoted as \(B^\star\). This additional
run follows the same overall training recipe and is included because it
provides stronger preparation-phase performance than the initial
DF-Arena checkpoint, especially for the speech component.

All base systems use the same component-level adaptation strategy. The
original classification layers of the pre-trained backbones are replaced
or bypassed by three binary classification heads,
$h_o(x)$, $h_s(x)$, and $h_e(x)$,
corresponding to the original/mixed, speech, and environment decisions.
Thus, all backbones are fine-tuned toward the same target representation,
which enables score-level fusion across heterogeneous architectures.

\subsection{Pre-Processing and Optimization}
All audio files are converted to mono waveforms and resampled to
16~kHz. The input waveform is deterministically cropped or repeat-padded
to the fixed input length required by the corresponding backbone. For
the DF-Arena implementation, this length is 64,600 samples, following
the original model-specific preprocessing setup.

The training objective is the weighted sum of three cross-entropy losses:
\begin{equation}
\mathcal{L}
=
\lambda_o \mathcal{L}_o
+
\mathbf{1}_{y \neq 0}
\left(
\lambda_s \mathcal{L}_s
+
\lambda_e \mathcal{L}_e
\right).
\end{equation}
where \(y \in \{0,\ldots,4\}\) is the five-class label and \(y=0\)
denotes original audio. The loss \(\mathcal{L}_o\) is computed for the
original/mixed head on all training samples. The speech and environment losses, \(\mathcal{L}_s\) and \(\mathcal{L}_e\), are computed only for mixed samples, since these component labels are defined only for non-original audio. We use loss weights
\(\lambda_o=\lambda_s=\lambda_e=1.0\). For the original/mixed head, we
use class weights \([0.2,0.8]\) for the mixed and original classes,
respectively.

All models are optimized using AdamW \cite{loshchilov2019decoupled}. The pre-trained backbone is fine-tuned with a learning rate of $10^{-5}$, while the newly initialized classification heads use a learning rate of $10^{-4}$. We employ a weight decay of $0.01$, a batch size of 32, and mixed-precision training, and train each model for 10 epochs. The learning rate is scheduled using ReduceLROnPlateau based on the validation macro-F1 score\footnote{\url{https://docs.pytorch.org/docs/stable/generated/torch.optim.lr_scheduler.ReduceLROnPlateau.html}}. All experiments are conducted on a single NVIDIA A100 GPU with 80~GB of memory.

\subsection{RawBoost Augmentation}
For each base architecture, we additionally train one variant with
RawBoost \cite{tak2022rawboostrawdataboosting}. RawBoost is applied
only during training with probability 0.5 and operates directly on the
waveform. We use the public RawBoost implementation with the setting
\texttt{algo=5}, corresponding to the LA configuration in that
implementation, which applies a combination of convolutive and impulsive
noise distortions\footnote{\url{https://github.com/TakHemlata/RawBoost-antispoofing/blob/main/data_utils_rawboost.py}}.
This augmentation introduces no external audio recordings or synthetic
samples. In our experiments, RawBoost is beneficial for some backbones
but not uniformly across all models, motivating the inclusion of both
augmented and non-augmented checkpoints in the ensemble search.

\subsection{Score Export}
For each trained checkpoint, we export a score file containing the three
binary head margins for every evaluation or test sample. Let
$z_{h,0}$ and $z_{h,1}$ denote the logits of head
$h \in \{o, s, e\}$. The exported margin is
\begin{equation}
d_h = z_{h,1} - z_{h,0}.
\end{equation}
For the original head, positive margins indicate higher confidence for
the original class. For the speech and environment heads, positive
margins indicate higher confidence for bona fide speech and bona fide
environment, respectively. These per-sample margins are used as inputs
to the ensemble.

\subsection{Ensemble Fusion}
All ensemble variants use margin-space fusion of component-level scores.
For each model $m$ and head $h \in \{o,s,e\}$, we use the exported
margin $d_h^{(m)}$ as the model score. The ensemble margin for each head
is computed as a weighted average:
\begin{equation}
\bar{d}_h = \sum_m \alpha_{h,m} d_h^{(m)}, \quad
\sum_m \alpha_{h,m} = 1.
\end{equation}
The fused margin is then converted to a probability-like score using the sigmoid function:
\begin{equation}
\bar{q}_h = \sigma\!\left(\bar{d}_h\right),
\end{equation}
where $\sigma(\cdot)$ denotes the sigmoid function. We use margin-space
fusion because the submitted component scores are logit margins and
preparation-phase ablations showed that directly averaging these margins
performed slightly better than probability-space fusion.

We evaluate two ensemble strategies: conservative fusion and
component-wise fusion. The conservative approach uses the same uniformly
weighted subset of models for all three heads. Conservative-4 uses the
set $\{A_{\mathrm{RB}}, B^\star, C_{\mathrm{RB}}, D\}$, while
Conservative-5 additionally includes $B$. In contrast, component-wise
fusion uses head-specific model weights, allowing each detector to
contribute primarily to the component for which it is most effective.
The corresponding weights are given in
Table~\ref{tab:specialist_weights}.

\begin{table}[t]
\centering
\caption{Final component-wise fusion recipe. Weights are normalized within
each component head. $B^\star$ denotes the additional no-RawBoost
DF-Arena run.}
\label{tab:specialist_weights}
\begin{tabular}{ll}
\hline
Head & Ensemble members and weights \\
\hline
Original &
$C_{\mathrm{RB}}: 0.70,\; A_{\mathrm{RB}}: 0.15,\; B_{\mathrm{RB}}: 0.15$ \\
Speech &
$B^\star: 0.959,\; B: 0.027,\; B_{\mathrm{RB}}: 0.014$ \\
Environment &
$C: 0.25,\; D_{\mathrm{RB}}: 0.25,\; A: 0.25,\; B^\star: 0.25$ \\
\hline
\end{tabular}
\end{table}

\subsection{Post-hoc Bias Calibration}
For the component-wise fusion only, we apply two lightweight post-hoc
calibration steps.

First, we add a constant bias to each fused component margin:
\begin{equation}
d'_h = \bar{d}_h + b_h .
\end{equation}
The head-specific biases are $b_o = -0.20$, $b_s = -0.30$, and
$b_e = -0.20$. The biased margins are then converted to component
probabilities using the sigmoid function:
\begin{equation}
q'_h = \sigma\!\left(d'_h\right).
\end{equation}
These calibrated component probabilities are mapped to the five-class
probabilities using the component factorization described above.

Second, we apply class-specific biases in log-probability space before
prediction:
\begin{equation}
\hat{y}
=
\arg\max_k
\left(
\log p_k + c_k
\right).
\end{equation}
We use $(c_0, c_1, c_2, c_3, c_4) =
(-0.12, 0.00, -0.15, -0.15, 0.10)$ as the class-bias vector.

The final submitted system corresponds to the component-wise ensemble
with both head-bias and class-bias calibration.

\section{Results and Analysis}
\label{sec:results}
This section evaluates the proposed component-level ensemble system on
the official ESDD2 evaluation and test sets. We first describe the
ranking and diagnostic metrics, then analyze the individual fine-tuned
models, the ensemble ablations, and the final comparison with the
official baseline.

\subsection{Evaluation Metrics}
The official ranking metric of the ESDD2 challenge is the macro-F1
score over the five target classes. We therefore use macro-F1 as the
primary metric throughout this section. In addition, we report the three
diagnostic equal error rates (EERs) provided by the challenge platform:
$\mathrm{EER}_{\mathrm{original}}$ for distinguishing original audio
from mixed audio, $\mathrm{EER}_{\mathrm{speech}}$ for detecting spoofed
speech components, and $\mathrm{EER}_{\mathrm{env}}$ for detecting
spoofed environmental components. Lower EER values indicate better
component-level discrimination, while higher macro-F1 indicates better
five-class classification performance.

\subsection{Single-Model Performance}
Table~\ref{tab:single_model_results} reports the official evaluation-set
performance of the individual fine-tuned models. All values are reported
as fractions. The results show that no single model dominates all
component-level metrics. For example, the additional DF-Arena run
$B^\star$ achieves the strongest single-model macro-F1, while other
models provide competitive or better scores for specific component EERs.
This heterogeneity motivates score-level fusion across models and
component heads.

\begin{table}[t]
\centering
\caption{Single-model performance on the evaluation set. RB denotes
RawBoost training. $B^\star$ is the additional no-RawBoost DF-Arena run
used in the ensemble.}
\label{tab:single_model_results}
\setlength{\tabcolsep}{4pt}
\begin{tabular}{lccccc}
\hline
Model & RB & Macro-F1 & EER$_o$ & EER$_s$ & EER$_e$ \\
\hline
$A$ (XLSR-Mamba) & no  & 0.7240 & 0.0193 & 0.3557 & 0.1540 \\
$A_{\mathrm{RB}}$ (XLSR-Mamba) & yes & 0.7265 & 0.0228 & 0.3110 & 0.1554 \\
\hline
$B$ (DF-Arena) & no  & 0.7184 & 0.0323 & \textbf{0.1816} & 0.2047 \\
$B_{\mathrm{RB}}$ (DF-Arena) & yes & 0.6974 & 0.0194 & 0.2274 & 0.1826 \\
$B^\star$ (DF-Arena) & no & \textbf{0.7350} & 0.0280 & 0.2111 & 0.1561 \\
\hline
$C$ (SLS) & no  & 0.6989 & 0.0179 & 0.4082 & \textbf{0.1182} \\
$C_{\mathrm{RB}}$ (SLS) & yes & 0.7227 & \textbf{0.0145} & 0.2740 & 0.3091 \\
\hline
$D$ (TCM-ADD) & no  & 0.7205 & 0.0359 & 0.2733 & 0.1494 \\
$D_{\mathrm{RB}}$ (TCM-ADD) & yes & 0.7098 & 0.0208 & 0.2949 & 0.1343 \\
\hline
\end{tabular}
\end{table}

RawBoost has a model-dependent effect. It improves the macro-F1 of
XLSR-Mamba and SLS, but not DF-Arena or TCM-ADD. Similarly, it can
improve one component metric while degrading another, as observed for
SLS, where RawBoost improves speech EER but worsens environmental EER.
Therefore, both RawBoost and non-RawBoost variants are retained as
candidate ensemble members rather than selecting a single augmentation
policy globally.

\subsection{Ensemble Ablation}
Table~\ref{tab:ensemble_results} compares the ensemble variants on the
evaluation and test sets. The conservative ensembles already improve
substantially over the individual models, confirming that score fusion
reduces model-specific errors. However, using the same model subset for
all heads is suboptimal. The component-wise fusion improves macro-F1 on
both evaluation and test sets, indicating that different models are
better suited to different component decisions.

\begin{table*}[t]
\centering
\caption{Official ensemble results on the evaluation and test sets.
Higher macro-F1 is better; lower EER is better.}
\label{tab:ensemble_results}
\setlength{\tabcolsep}{4pt}
\begin{tabular}{lcccccccc}
\hline
\multirow{2}{*}{System}
& \multicolumn{4}{c}{Evaluation set}
& \multicolumn{4}{c}{Test set} \\
\cline{2-5} \cline{6-9}
& Macro-F1 & EER$_o$ & EER$_s$ & EER$_e$
& Macro-F1 & EER$_o$ & EER$_s$ & EER$_e$ \\
\hline
Conservative-4
& 0.7555 & 0.0138 & 0.2289 & 0.1463
& 0.7649 & 0.0122 & 0.2263 & 0.1362 \\
Conservative-5
& 0.7551 & 0.0147 & \textbf{0.2125} & 0.1488
& 0.7667 & 0.0125 & \textbf{0.2073} & 0.1373 \\
Component-wise fusion
& 0.7649 & 0.0124 & 0.2177 & 0.1341
& 0.7766 & 0.0109 & 0.2164 & 0.1263 \\
+ head bias
& 0.7710 & 0.0124 & 0.2177 & 0.1341
& 0.7803 & 0.0109 & 0.2164 & 0.1263 \\
+ head and class bias
& \textbf{0.7715} & \textbf{0.0124} & 0.2177 & \textbf{0.1341}
& \textbf{0.7828} & \textbf{0.0109} & 0.2164 & \textbf{0.1263} \\
\hline
\end{tabular}
\end{table*}

The strongest configuration is the component-wise fusion with both
head-bias and class-bias calibration. Head-bias calibration provides the
largest improvement over the uncalibrated component-wise ensemble,
increasing macro-F1 from 0.7649 to 0.7710 on the evaluation set and from
0.7766 to 0.7803 on the test set. Adding class-bias calibration gives a
smaller but consistent further gain, reaching 0.7715 macro-F1 on the
evaluation set and 0.7828 macro-F1 on the test set.

The ranking of the ensemble variants is consistent across the evaluation
and test sets: component-wise fusion with head and class bias performs
best, followed by head-bias-only calibration, uncalibrated component-wise fusion, and the conservative ensembles. The only minor exception is that Conservative-4 slightly outperforms Conservative-5 on the evaluation set, whereas Conservative-5 is better on the test set. This suggests that adding an additional model can improve generalization.

EER depends on the relative ordering of the component scores rather than
on one fixed decision threshold. Thus, adding a constant head bias shifts
all scores equally and leaves the component EERs unchanged. Class-bias
calibration is used only for the final five-class argmax decision, so it
can affect macro-F1 but not the component EERs.

\subsection{Leaderboard Comparison}
Table~\ref{tab:leaderboard_comparison} shows the leaderboard context for
both challenge phases. Our submission ranked 5th in the final ranking
phase with 0.7828 macro-F1 on the hidden test set and 3rd in the
preparation phase with 0.7715 macro-F1 on the evaluation set. Compared
with the official baseline, this corresponds to an absolute macro-F1
improvement of 0.1501 on the test set and 0.1491 on the evaluation set.

\begin{table}[t]
\centering
\caption{Leaderboard context for the final ranking and preparation
phases. The official baseline is included for reference.}
\label{tab:leaderboard_comparison}
\setlength{\tabcolsep}{4pt}
\begin{tabular}{lcccc}
\hline
Participant & Test rank & Test F1 & Eval rank & Eval F1 \\
\hline
orange9 & 1 & 0.8775 & 2 & 0.8029 \\
xiaoxuan\_guo & 2 & 0.8266 & 5 & 0.7647 \\
seantang & 3 & 0.8200 & 1 & 0.8124 \\
sv4g\_team & 4 & 0.8077 & 13 & 0.7120 \\
\textbf{Ours} & \textbf{5} & \textbf{0.7828} & \textbf{3} & \textbf{0.7715} \\
temmiexpratt & -- & -- & 4 & 0.7653 \\
\hline
Official baseline & 26 & 0.6327 & 33 & 0.6224 \\
\hline
\end{tabular}
\end{table}

The official baseline uses a separation-enhanced joint learning
framework: a global detector first identifies potentially spoofed
mixtures, a separation module estimates speech and environmental
components, and component-specific detectors are then fused into the
five-class decision. In contrast, our system performs component-level
prediction directly from the input waveform and relies on multi-backbone
margin-space score fusion. For our final submission, the diagnostic EERs
are 0.0109, 0.2164, and 0.1263 for the original, speech, and environment
components, respectively. Compared with the official baseline
(0.0173/0.1978/0.4279), the largest diagnostic improvement is observed
for environmental sound detection. The speech EER is slightly worse than
the official baseline, but the overall macro-F1 is substantially higher,
indicating that improved five-class decision calibration and stronger
original/environment decisions outweigh this diagnostic trade-off. Without released hidden metadata, we cannot attribute this to specific
attack sources.

The evaluation-to-test change of our system is comparable to that of the
official baseline, while several other submissions show larger gains in
the final ranking phase. This may reflect additional phase-specific
system updates during the test phase.

\subsection{Discussion}
The results support three main observations. First, validation-set
performance alone is not sufficient for reliable model selection in this
challenge. This is consistent with the dataset design: the training and
validation splits share one source distribution, while the evaluation and
test splits share another. Second, RawBoost is useful but not uniformly
beneficial; its effect depends on both the backbone and the component
head. Third, component-wise fusion is more effective than uniform
conservative fusion, because the best detector for one component is not
necessarily the best detector for another. The final head and class
biases further improve macro-F1 by adjusting decision thresholds and
class priors after fusion.

The main cost of the proposed system is inference complexity. The final ensemble evaluates eight checkpoint instances per clip,
corresponding to approximately 4.6B backbone parameters. Compared with
the strongest single model on the evaluation set, \(B^\star\), the
ensemble improves macro-F1 from 0.7350 to 0.7715, but requires multiple
forward passes.

\section{Conclusion}
\label{sec:conclusion}
We presented a component-level ensemble system for the ESDD2 speech and
environmental sound deepfake detection challenge. Instead of directly
training a single five-class classifier, we factorized the task into
three binary decisions: original versus mixed audio, bona fide versus
spoofed speech, and bona fide versus spoofed environmental sound. Four
pre-trained anti-spoofing models were fine-tuned using this common
formulation, with additional RawBoost-augmented variants for each
backbone. The results show that individual models exhibit complementary
strengths across the three component heads, motivating score-level
fusion.

Our best system uses margin-space component-wise fusion followed by
head-bias and class-bias calibration. This configuration consistently
outperformed conservative uniform ensembles on both the evaluation and
test sets. It achieved 0.7715 macro-F1 on the evaluation set, ranking
3rd in the preparation phase, and 0.7828 macro-F1 on the hidden test
set, ranking 5th in the final ranking phase. The results also highlight
that validation performance was not always a reliable proxy for
evaluation and test performance, which is consistent with the
distribution shift between the development and leaderboard splits.
Overall, the proposed system demonstrates that component-aware
multi-backbone fusion is an effective strategy for composite audio
deepfake detection.

\section*{Acknowledgment}
This research was supported through the National Research Center for Applied Cybersecurity ATHENE. ATHENE is funded by
the Federal Ministry of Education and Research and the Hessian Ministry of Higher Education, Research, Science and the Arts.

\bibliographystyle{IEEEbib}
\bibliography{bib}

\end{document}